\documentclass[sigconf]{acmart}

\usepackage{bm} 
\usepackage{xfrac} 
\usepackage{booktabs} 
\usepackage{algorithm}
\usepackage[noend]{algpseudocode}
\usepackage{caption}
\usepackage[margin=20pt]{subcaption}
\usepackage{hyperref}
\usepackage{enumerate}
\usepackage{mymacros}
\usepackage{balance} 
\usepackage{paralist}
\usepackage{dsfont}
\usepackage{rotating}
\usepackage{listings}
\usepackage{adjustbox}

\usepackage[subtle]{savetrees}

\usepackage[font={small},skip=0.5pt]{caption}
\usepackage{setspace}

\usepackage{graphicx}
\usepackage[normalem]{ulem}
\useunder{\uline}{\ul}{}

\renewcommand\footnotetextcopyrightpermission[1]{} 

\begin{document}

\title{Zero to Hero: Exploiting Null Effects to Achieve Variance Reduction in Experiments with One-sided Triggering}
\author{Alex Deng}
\affiliation{
  \institution{Airbnb}
  \city{Seattle} 
  \state{WA} 
  \country{USA}
  \postcode{98101}
}
\email{alex.deng@airbnb.com}

\author{Lo-Hua Yuan}
\affiliation{
  \institution{Airbnb}
  \city{San Francisco} 
  \state{CA} 
    \country{USA}
  \postcode{94103}
}
\email{lohua.yuan@airbnb.com}

\author{Naoya Kanai}
\affiliation{
  \institution{Airbnb}
  \city{San Francisco} 
  \state{CA} 
    \country{USA}
  \postcode{94103}
}
\email{naoya.kanai@airbnb.com}

\author{Alexandre Salama-Manteau}
\affiliation{
  \institution{Airbnb}
  \city{San Francisco} 
  \state{CA} 
    \country{USA}
  \postcode{94103}
}
\email{alexandre.salama-manteau@airbnb.com}

\renewcommand{\shortauthors}{Alex Deng, Lo-Hua Yuan, Naoya Kanai, \& Alexandre Salama-Manteau}


\settopmatter{printfolios=false}

\pagestyle{plain}

\begin{abstract}
In online experiments where the intervention is only exposed, or ``triggered'', for a small subset of the population, it is critical to use variance reduction techniques to estimate treatment effects with sufficient precision to inform business decisions.
Trigger-dilute analysis is often used in these situations, and reduces the sampling variance of overall intent-to-treat (ITT) effects by an order of magnitude equal to the inverse of the triggering rate; for example, a triggering rate of $5\%$ corresponds to roughly a $20x$ reduction in variance.
To apply trigger-dilute analysis, one needs to know experimental subjects' triggering counterfactual statuses, i.e., the counterfactual behavior of subjects under both treatment and control conditions. 
In this paper, we propose an unbiased ITT estimator with reduced variance applicable for experiments where the triggering counterfactual status is only observed in the treatment group. 
Our method is based on the efficiency augmentation idea of CUPED and draws upon identification frameworks from the principal stratification and instrumental variables literature. 
The unbiasedness of our estimation approach relies on a testable assumption that the augmentation term used for covariate adjustment equals zero in expectation. 
Unlike traditional covariate adjustment or principal score modeling approaches, our estimator can incorporate both \emph{pre-experiment} and \emph{in-experiment} observations. 
We demonstrate through a real-world experiment and simulations that our estimator can remain unbiased and achieve precision improvements as large as if triggering status were fully observed, and in some cases can even outperform trigger-dilute analysis. 
\end{abstract}


\maketitle

\section{Introduction}\label{sec:intro}
Web search and online user-facing products have a long history of using online controlled experiments, also known as A/B tests, in order to measure the impact of new features and designs and to accelerate product iteration~\citep{Gupta:2019,kohavi2009controlled,googlesurvey,xu2015infrastructure,xie2021measure,bakshy2014designing,Deng:2016b,drutsa2019effective}.
In online A/B testing, treatment assignment typically occurs upstream of treatment exposure, and experimental subjects must first trigger an event in order to be exposed to the intervention of interest.
For example, subjects have to open a promotion email before they are exposed to one of multiple discount promotion deals.
In another instance, subjects have to query a particular keyword before they are shown one of multiple sets of search results.

\begin{figure*}[bthp!]
    \centering
    \includegraphics[clip, trim=0.01cm 1.3cm 0.1cm 3.3cm, width=0.9\textwidth]{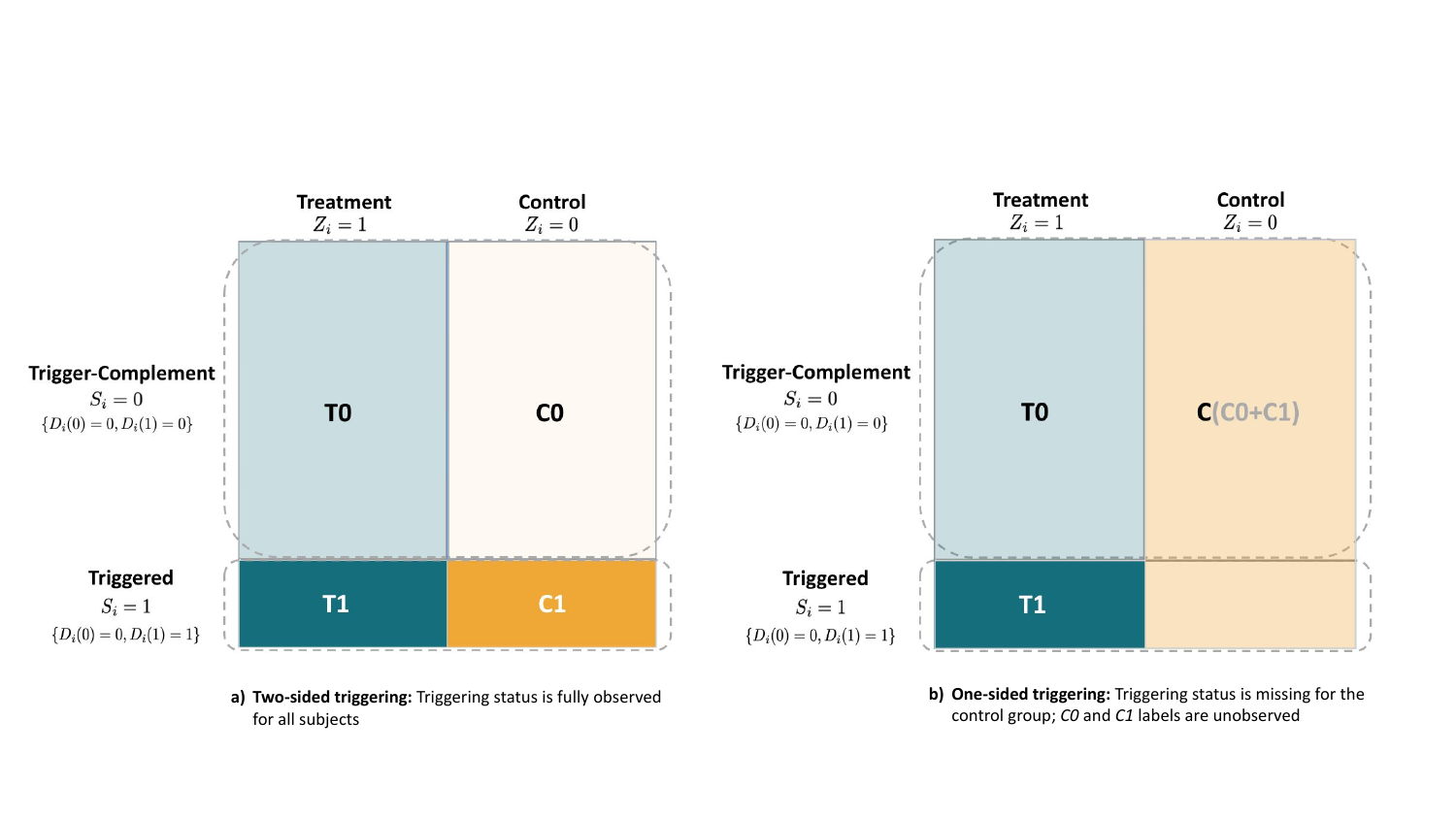}
    \caption{Illustration of (a) two-sided triggering and (b) one-sided triggering. $T1$ are subjects assigned to treatment who triggered exposure to the active treatment. $C1$ are subjects assigned to control who triggered exposure to the control condition (and who would have triggered exposure to the active treatment if assigned to the treatment group). $T0$ and $C0$ are called the trigger-complement group and consist of subjects who would not trigger exposure to the active treatment regardless of which study group they are assigned to. In one-sided triggering, $C1$ and $C0$ labels are unobserved.}
    \label{fig:trigger_illustrate}
\end{figure*}

The presence of triggering defines four sub-populations of interest, illustrated in the left panel of Figure~\ref{fig:trigger_illustrate}: 
In the treatment group, there are those who triggered exposure to the active treatment condition ($T1$) and those who did not trigger ($T0$). 
Similarly, in the control group, there are those who triggered exposure to the control condition (meaning they would have triggered exposure to the active treatment had they been assigned to the treatment group) ($C1$), and those who did not trigger and would not have triggered even if assigned to treatment ($C0$). 
In an experiment with triggering conditions, the naive difference-in-means between treatment and control outcomes is an unbiased but non-optimal estimator for the overall intent-to-treat effect. 
A more precise estimator can be obtained via trigger-dilute analysis: estimate an average treatment effect for the triggered subjects ($T1$ vs. $C1$) and then multiply this triggered average treatment effect by the triggering rate (also called coverage).
As a rough heuristic, trigger-dilute analysis can reduce sampling variance by a factor of the inverse of the triggering rate, e.g., approximately $20x$ variance reduction with a triggering rate of $5\%$. 
Such a massive variance reduction is too valuable to overlook \cite{kohavi2020trustworthy}.

In this paper, we study the case where triggering status is not observed for control subjects, i.e., we do not observe the $C1$ and $C0$ triggering counterfactual labels, so trigger-dilute analysis cannot be used.
This one-sided triggering scenario, shown in Figure~\ref{fig:trigger_illustrate}(b), is known as one-sided noncompliance in the causal inference literature.  
(See Table~\ref{tab:notation} for a mapping of terminology between the A/B testing and causal inference literature.) 
Such cases naturally arise for interventions that require users to opt in to a new feature or experience.
For example, users are sent an invitation to opt in to a new program, but their behavior will not be directly affected unless they accept the invitation. 
Alternatively, a navigation button is added for a new functionality, but users will not experience the new functionality unless they click on the button.\footnote{Sending invitations or adding navigation buttons may induce secondary effects unrelated to the new program or functionality. These secondary effects are not of research interest in this paper, but can be tracked and investigated through diagnostic metrics.}  

For one-sided triggered experiments, we propose an estimator that is unbiased for the overall intent-to-treat (ITT) effect and has smaller variance than the default difference-in-means estimator $\dd{Y} := \wbar{Y}_T - \wbar{Y}_C$. 
We take a model-assisted rather than model-based estimation approach, and build upon the idea of CUPED~\cite{deng2013cuped}, which is to engineer a mean-zero augmentation term $\that_0$ by which we can adjust the unbiased $\dd{Y}$ estimator, e.g., use $\dd{Y} - \theta\cdot \that_0$, for some smartly-chosen value $\theta$.  
The key assumption is that there is no treatment effect for $T0$ and $C0$ subjects; these are trigger-complement subjects who would not be exposed to the intervention regardless of their study group assignment. 
We can then use observations from $T0$ and $C0$ to construct $\that_0$.
Since trigger-complements are a subset of the full experimental population, $\that_0$ and $\dd{Y}$ will be correlated, and strongly so when the triggering rate is low and trigger-complements comprise a larger proportion of the full experiment population.
By the theory behind CUPED, this (strong) correlation between $\that_0$ and $\dd{Y}$ will provide (potentially massive) variance reduction.

\begin{table}[htbp!]
\resizebox{0.9\columnwidth}{!}{%
\begin{tabular}{@{}ll@{}}
\toprule
A/B Testing                                 & Causal Inference                                                                                   \\ \midrule
Treatment Assignment                        & Instrumental Variable                                                                                \\
Triggering Counterfactual Status                  & Compliance Status (Principal Stratum)                                                                                         \\
Only Triggered subjects can be affected by the treatment     & Exclusion Restriction                                                                              \\
Only Treatment group can receive active treatment  &  Strong Monotonicity  \\
Triggered Average Treatment Effect          & \begin{tabular}[t]{@{}l@{}}Local Average Treatment Effect \\ (Complier Average Treatment Effect)\end{tabular} \\
Overall Average Treatment Effect (Intent-to-Treat Effect)          & Intent-to-Treat Effect                                                                           \\ 
T1, C1 (Triggered)           & Compliers                                                                             \\ 
T0, C0 (Trigger-Complement)          & Never-Takers                                                                             \\ 
Triggering Probability               & Principal Score \\
\bottomrule
\end{tabular}%
}
\caption{Terminology mapping between A/B testing and the causal inference literature. We focus on the typical case for online experiments, where by design there are no Always-Takers nor Defiers because the active treatment experience is restricted to those assigned to the treatment group. Thus, the Triggered Average Treatment Effect is equivalent to the Local (Complier) Average Treatment Effect.}
\label{tab:notation}
\end{table}

\subsection{Setup and Notation}\label{sec:notation}
We consider a randomized experiment with binary treatment assignment $Z_i \in \{0,1\}$.
A binary label $D_{i}\left(Z_i=z\right) \in \{0, 1\}$ indicates whether a subject triggered exposure to the active treatment when assigned to study group $z$. 
For example, $D_{i}(0) = 1$ indicates that subject $i$ was assigned to the control group and triggered exposure to the active treatment.
$\{Y_i(0), \ Y_i(1)\}$ is the pair of counterfactual outcomes under assignment to control and treatment, respectively~\cite{Imbens:2015}.
Similarly, $S_i = \{D_i(0), \; D_i(1)\}$ is the counterfactual pair of triggered exposure conditions under assignment to control and treatment~\cite{angrist1994,frangakis2002}.
We call $S$ the ``triggering counterfactual status'', or ``triggering status'', for short.

We focus on the typical case for online experiments where $D_i(0) = 0$ by design; that is, subjects assigned to control have no access to the active treatment and can only be exposed to the control condition.
This allows us to simplify notation: We use $S_i = 1$ interchangeably with $S_i=(D_i(0)=0, \; D_i(1)=1)$ to denote subjects who would trigger exposure to the active treatment if and only if assigned to the treatment group ($T1$ and $C1$).
We use $S_i = 0$ interchangeably with $S_i=(D_i(0)=0,\; D_i(1)=0)$ to denote subjects who would not trigger exposure to the active treatment, regardless of their treatment assignment ($T0$ and $C0$).
Mapping to Figure~\ref{fig:trigger_illustrate}, $S_i=1$ corresponds to the Triggered group while $S_i=0$ corresponds to the Trigger-Complement group.
$Y_i$ and $D_i$ refer to subject $i$'s observed outcome and observed exposure condition, respectively.
We use subscripts $T$ and $C$ to denote treatment and control groups, so $\Delta(Y) := \overline{Y}_T - \overline{Y}_C$ denotes the ``delta'' of the average observed outcome $Y$ between the two study groups.
The causal impact we wish to estimate is the overall ITT effect $\EE[Y_i(1) - Y_i(0)]$, also known as the overall average treatment effect.

\subsection{Related Work}\label{sec:related}
In the online A/B testing literature, trigger-dilute analysis is a popular approach for obtaining a more precise ITT estimate when the intervention is only exposed to a small subset of the experimental population~\cite{kohavi2007practical,deng2015diluted,abScale}.
First, an average treatment effect is estimated among the triggered subjects.
This triggered average treatment effect is then multiplied (``diluted'') by the triggering rate. 
The success of a trigger-dilute analysis depends on either a simple triggering condition, such as a user visiting a specific webpage where the experience differs, or a mechanism of counterfactual logging such that the experimentation system is able to compare the treatment vs. control experience at any time to label whether a realized difference exists between the study groups' respective experiences. 
Complex triggering conditions often lead to sample ratio mismatch, rendering the triggered analysis untrustworthy~\cite{fabijan2019diagnosing}. 
Moreover, when user opt-in is the triggering signal, triggering status is by definition not observed for control units. 
When triggering status is not observed for all units, trigger-dilute analysis cannot be applied.

To tackle scenarios with partially-observed triggering, we turn to the causal inference literature around noncompliance, which occurs when treatment assignment differs from treatment exposure (i.e., $Z_i \neq D_i$ for some units $i$). 
Instrumental variables (IV)~\cite{angrist1994, angrist1996} and principal stratification~\cite{frangakis2002} frameworks provide strategies for identifying and estimating subgroup average treatment effects under a range of noncompliance conditions. 

In the IV literature, the causal quantity of interest is typically the local average treatment effect (LATE), i.e., the intervention effect among Compliers.
Since here we focus on the typical online testing scenario where exposure to the active treatment is restricted to those who are assigned to the treatment group (so there are no Always-Takers nor Defiers), LATE is equivalent to the Triggered Average Treatment Effect.
The standard IV estimator for LATE is $\Delta(Y)$ divided by the estimated proportion of Compliers (i.e., the triggering rate). 
Multiplying this IV estimator by the triggering rate therefore recovers $\Delta(Y)$. 
As such, the standard IV approach is not aimed at producing a more precise estimate of the overall ITT.
There are, however, extensions of the standard IV estimator that attempt to improve estimation efficiency.
For example, weighted IV methods~\cite{Coussens2021ImprovingIF, HuntingtonKlein_2020, joffe2003weighting} use predicted compliance to weight both treatment and control groups when computing the LATE estimator. 
Our method is similar in that we also use predicted compliance, but whereas the weighted IV estimator is unbiased for LATE only when there is no correlation between treatment effect heterogeneity and compliance, our estimator is unbiased for ITT (and asymptotically unbiased for LATE after dividing by the triggering rate) as long as the augmentation term that we construct equals zero in expectation.
Also, unlike weighted IV, our method can utilize in-experiment observations to predict compliance probabilities.

Beyond instrumental variables, the principal stratification literature generalizes identification and estimation strategies for causal effects under more complicated forms of noncompliance~\cite{ding2017principal,feller2017,yuan2019,jiang2020identification}.
We draw upon many ideas from the principal stratification literature to construct our proposed estimator.
In particular, we invoke a key assumption called \emph{weak principal ignorability}~\cite{jo_stuart2009,feller2017,ding2017principal} as a sufficient condition under which our estimator is unbiased for ITT.
We also use principal scores (i.e., predicted triggering probabilities) to construct the augmentation term critical to our estimator. 

More generally, there is a vast literature on using pre-experiment covariates for regression adjustment to increase estimation efficiency~\cite{fisher1925statistical,guo2021machine,poyarkov2016boosted,xie2016improving,lin2013agnostic,liding2020}. 
Our method is based on the augmentation idea of CUPED~\cite{deng2013cuped}(see also~\cite{liding2020}), applied to the one-sided triggering context, and is a general approach that can be used on top of any pre-experiment covariate regression adjustment. 
One unique advantage of our approach is its flexibility to incorporate in-experiment observations without introducing bias.

\subsection{Contribution and Organization}\label{sec:contribution}
This paper makes the following contributions to experimentation and causal measurement:
\begin{enumerate}
    \item We propose an unbiased ITT estimator with reduced variance for experiments with one-sided triggering. Our estimator relies on a testable assumption that an augmentation term used for covariate adjustment equals zero in expectation.
    \item We explain how to test for this mean-zero assumption. When the augmentation term fails a mean-zero test, we show how our estimator can incorporate in-experiment observations to reduce or eliminate the augmentation's bias. This idea is novel and effective for many real applications. 
    \item We demonstrate our method through a real experiment where using both \emph{pre-experiment} and \emph{in-experiment} covariates reduced variance by a massive factor of 60. We study multiple flavors of augmentation and explain their differences in theory and through simulation studies.
    \item Our method is straightforward to implement. We provide code\footnote{\url{https://osf.io/kum6d/?view_only=8c2f0fe6c40c40029faecb7a583f0c45}} to reproduce the simulation study.
\end{enumerate}

\noindent The rest of the paper is organized as follows: 
Section~\ref{sec:cuped} 
starts with a review of CUPED and proposes an augmentation-based estimator for experiments with one-sided triggering.
Section~\ref{sec:cov_bal} discusses ways to test the mean-zero assumption that is crucial for the unbiasedness of our proposed estimator. 
We offer guidance on how to select covariates to construct the augmentation term, and explain how using in-experiment data to achieve the mean-zero condition relates to a bias-variance tradeoff.
We illustrate our method with applications to a real experiment in Section~\ref{sec:realexp} and a series of simulation studies in Section~\ref{sec:simulation}. 
Section 6 concludes and addresses limitations of our approach.

\section{Variance Reduction using CUPED}\label{sec:cuped}
CUPED, acronym for Controlled-experiment Using Pre-experiment Data~\cite{deng2013cuped}, is a variance reduction technique widely adopted in the A/B testing industry to improve the sensitivity of A/B tests~\cite{xie2016improving, kohavi2020trustworthy,poyarkov2016boosted}. 
At its core, CUPED is an efficiency augmentation method applied on top of any existing unbiased estimator. 
If $\that$ is an unbiased estimator for $\tau$, the basic CUPED estimator is defined as 
\begin{equation}
    \that_{cuped.basic}~:=~\that - \that_0 \ ,  \label{eq:cuped}
\end{equation}
where $\that_0$ is an augmentation such that $\EE (\that_0)=0$. The mean-zero requirement ensures that the CUPED estimator has the same expectation as the original estimator $\that$, and therefore is also unbiased for $\tau$. 
Variance reduction is achieved when there is sufficient correlation between $\that$ and $\that_0$, since
$$
\var (\that - \that_0) = \var (\that) + \var (\that_0) - 2\cov (\that, \that_0) < \var (\that)
$$
whenever $\var (\that_0) < 2\cov (\that, \that_0)$.  
Moreover, for any fixed $\theta$, $\theta \cdot \that_{0}$ is also a mean-zero augmentation, so we can solve for a $\theta$ that minimizes the variance of the new augmentation 
\begin{equation}
    \that_{cuped}: = \that - \theta\cdot \that_{0}\ .  \label{eq:cuped2}
\end{equation}
\citet{deng2013cuped} showed that this optimization has a solution similar to an ordinary least squares regression, where $\theta = \cov(\that, \that_{0})/\var(\that_{0})$ is the multiplier that minimizes the variance of $\that_{cuped}$, providing a variance reduction rate equal to $\cor(\that,\that_{0})^2$.

\subsection{CUPED for One-Sided Triggering}\label{sec:cuped-oneside}
To apply CUPED on experiments with one-sided triggering, we search for constraints in the data generating process that can be used to construct mean-zero augmentations.
We exploit a key assertion that there is no treatment effect for the trigger-complement group, e.g., the intervention has no effect on the $T0$ and $C0$ subjects who would never opt-in regardless of whether we give them the choice to opt-in.
This means that if we can transform the entire control group $C (C0 + C1)$ to make it comparable to $T0$ in terms of the outcome $Y$ distribution, then we can use the difference in outcomes between $T0$ and transformed $C$ for mean-zero augmentation. 
And since $T0$ and $C$ are random samples from the full experimental population, we can guarantee correlation between this augmentation term and $\dd{Y}$, thereby obtaining a more precise ITT estimate, following Eq~(\ref{eq:cuped2}).

To make $T0$ and $C$ comparable, we leverage matching and covariate balancing techniques from the causal inference literature~\cite{Imbens:2015,rosenbaum1983central,chattopandhyay2022}. 
The trick is to properly reweight subjects in $C$ by their probability of being among $C0$. 
Under randomization, the triggering rates in the treatment and control groups are equal in expectation, so we can use the treatment group to fit a triggering probability model~\cite{jo_stuart2009,ding2017principal,feller2017}, and apply this model to $C$ to estimate each control subject's probability of belonging to $C0$.
Given a set of weights $w_i$ for each data point in the control group $C$, we can define the augmentation term
\begin{equation}\label{eq:h_sn}
\that_{0} :=    \frac{\sum_{i \in T}((1-D_i) Y_i)}{\sum_{i \in T} (1- D_i)} - \frac{\sum_{i \in C}(w_i Y_i)}{\sum_{i \in C} w_i} \ .
\end{equation}
The zero subscript in $\that_{0}$ reminds us that we are trying to identify the $T0$ and $C0$ trigger-complement subjects.
$\that_{0}$ is the difference of two weighted averages, where for treated subjects we use a hard weight of observed $D_i$, and for control subjects we use a soft weight $w_i$.\footnote{Compliance weighted IV estimators use soft weights $w_i$ for both treatment and control units. This won't provide a mean-zero augmentation in our framework because $Y$ from $T1$ will carry a treatment effect to bias our augmentation.} 
The first term on the right hand side of Eq~\eqref{eq:h_sn} is just $\overline{Y}_{T0}$, the sample average of $Y$ in $T0$.
The second term is an importance sampled average of the control subjects to mimic the average of $Y$ in $C0$. 

\subsubsection{\bf Finding appropriate weights}\label{sec:weights}
How do we construct the weights $w_i$ for the control group such that $\EE {\left[\that_{0}\right]} = 0$? We can take either a "prediction approach" or a "balancing approach".

The \textbf{prediction} approach uses a set of pre-experiment covariates $\bX$ to predict the triggering probability, or \emph{principal score}, of each subject $i$, and defines $w_i = 1 - \PP(S_i=1|\bX_i)$.
The augmentation~\eqref{eq:h_sn} will then be mean-zero under \emph{weak principal ignorability}~\citep{feller2017, jiang2020identification}.
Intuitively, this ignorability assumption states that, for individuals assigned to control, whether they \emph{would have triggered the active treatment if offered} is unrelated to their outcome given covariates.
Here, we formally state the assumption and theorem, and provide a proof in the online Appendix\footnote{ \url{https://osf.io/dfau5/?view_only=d9814c3ea8ba4314949f93c8f5cc241d}}:
\begin{ass}[Weak Principal Ignorability]\label{ass:ignorability}
There exists a set of pre-experiment covariates $\bX$ such that
$$
\EE (Y_i(0)|S_i=1, \bX_i) = \EE (Y_i(0)|S_i=0, \bX_i) = \EE (Y_i(0)|\bX_i) \ .
$$
This is implied by a slightly stronger assumption that $Y_i(0)$ and the triggering counterfactual $S_i$ are conditionally independent given $\bX_i$.
\end{ass}

\begin{thm}\label{thm:zero-mean-one-side-aug}
Under weak principal ignorability Assumption~\eqref{ass:ignorability}, the expectation of~\eqref{eq:h_sn} asymptotically equals $0$.
\end{thm}

The \textbf{balancing} approach, in contrast, aims to find weights that directly balance $T0$ and $C$. 
One option is to use the balancing property of the \emph{propensity score} $e(\bX):=\PP(T0|\bX, T0 \cup C)$, and define $w_i = e(\bX_i)/\left(1 - e(\bX_i)\right)$. 
(We refer readers to the rich propensity score literature~\cite{rosenbaum1983central, Imbens:2015} for assumptions and proof of the mean-zero claim under this balancing score.)
Another procedure is to directly solve for $w_i$ such that a rich set of covariates $\bX$ is almost perfectly balanced between $T0$ and the reweighted $C$, as studied extensively in the covariate balancing literature~\cite{imai2014covariate, hainmueller2012entropy, qingyuan2017entropy}.  

A major advantage of the balancing approach over the prediction approach is that the direct balancing approach can include \emph{in-experiment} observations in the set of covariates $\bX$. 
This inclusion of in-experiment covariates is a novel aspect of our method. 
This is possible because under the framework of one-sided triggering, neither subjects in $T0$ nor $C$ are exposed to the active treatment. 
Hence, all in-experiment observations for $T0$ and $C$ are free of a treatment effect and can be used for balancing.
We can even go to the extreme of directly balancing the target outcome $Y$, forcing~\eqref{eq:h_sn} to equal 0 exactly. 
But as we later explain, doing so will undermine variance reduction. 
It is generally preferred to use pre-experiment covariates, and to only include in-experiment covariates when necessary, i.e., when relying solely on pre-experiment covariates cannot create an augmentation~\eqref{eq:h_sn} that passes a mean-zero test. 
We discuss this further in Section~\ref{sec:cov_bal}. 

\subsection{\bf One-Sided Trigger Estimator}\label{sec:oneside-trigger-steps} 
\noindent To apply the $\that_{0}$ mean-zero augmentation with CUPED, we carry out the following steps:
\begin{enumerate}
    \item Find weights $w_i$ using the prediction or balancing approach. 
    \item Define $\that_{0}$ as in Eq~\eqref{eq:h_sn} and
    define $\theta = \frac{\cov(\dd{Y},~\that_{0})}{ \var (\that_{0})}$. \
    \item The CUPED One-Sided Trigger estimator is 
    \begin{equation}\label{eq:mutrig}
        \widehat{\tau}_{trg1}~:=~\Delta(Y) ~ - ~\theta \cdot \that_{0} \ ,
    \end{equation}
        and has variance
    \begin{equation}\label{eq:mutrig-var}
        \var(\widehat{\tau}_{trg1})~=~\var(\Delta(Y))~-~ \cov(\Delta(Y),~\that_{0})^2~/~\var(\that_{0}).
    \end{equation}
\end{enumerate}
We use bootstrap~\citep{efron1994introduction} to estimate the variances of $\Delta(Y)$, $\hat{\tau}_0$, and their covariance, and recalculate the weights $w_i$ for each bootstrap sample.
$\theta$ is computed using bootstrapped covariance and variance.

\section{Testing the Mean-Zero Assumption: a Bias-Variance Tradeoff}\label{sec:cov_bal}
In Section~\ref{sec:cuped-oneside}, we discussed how $\widehat{\tau}_{trg1}$ is asymptotically unbiased for ITT under weak principal ignorability.
This ignorability condition is not testable because it is based on counterfactuals $Y_i(0)$ and $S_i$ which are not observable at the same time.
However, we can still make practical progress by way of the testable assertion that $\tghat$ is unbiased for ITT as long as the augmentation $\that_0$ equals zero in expectation. 
Examples include using a Wald test, the delta method~\cite{Dengkdd2018} to compute the variance of $\that_0$, or bootstrapped $\widehat{\var}(\that_0)$, as described in Section~\ref{sec:oneside-trigger-steps}.

Only requiring $\EE(\that_0)=0$ means practitioners have a lot of freedom to modify and improve upon the $\that_{0}$ augmentation term~\eqref{eq:h_sn}.
As is common in observational data analyses, we can apply procedures such as weight bucketing, weight trimming, outcome outlier removal, etc., to further centralize~\eqref{eq:h_sn} towards a mean of 0. 
Additionally, the balancing approach can use \emph{in-experiment} observations.
In one extreme case, by treating the outcome of interest $Y$ itself as an in-experiment covariate, we can force~\eqref{eq:h_sn} to be exactly 0 and trivially satisfy the mean-zero requirement. 
However, this would also forgo any precision gains.
Following Eq~\eqref{eq:mutrig-var}, if $\that_{0}$ is a point mass at zero, then $\var(\tghat) = \var\left(\dd{Y}\right) - 0 = \var\left(\dd{Y}\right)$.

In theory, we strive to find the minimum set of covariates satisfying the weak ignorability Assumption~\eqref{ass:ignorability}. 
In practice, when pre-experiment covariates do not capture all relevant confounders of triggering status and outcome, balancing on in-experiment covariates can pull $\that_{0}$ closer to zero, in exchange for balancing away ``good'' variation captured by the augmentation term, thereby weakening efficiency gains. 
This is an interesting and novel observation with important practical implications: 
\begin{enumerate}
    \item We are guaranteed to have a mean-zero augmentation by balancing $Y$ directly between $T0$ and $C$, but with no variance reduction.
    \item We can start by balancing on a small set of pre-experiment covariates to construct an augmentation that reduces variance. But this augmentation may fail to pass the mean-zero test if we are missing important confounders of triggering status and outcome.
    \item We can gradually add more covariates for balancing, including in-experiment observations that are highly correlated with $Y$. Adding more balancing covariates can de-bias the augmentation, but will also lessen the amount of variance reduced.
\end{enumerate}

\subsection{\bf Which Covariates to Adjust or Balance?}\label{sec:adjust-or-not} 
Consider the law of total variance:
$\var(Y) = \var \left \{ \EE(Y|\bX) \right \} + \EE \left \{ \var(Y|\bX) \right \}\ ,
$
in which we decompose the variance of $Y$ into the variance explained by $\bX$
and the remaining variance not explained by $\bX$. 
In analyses of randomized experiments, removing variance attributed to empirical imbalances in pre-experiment covariates generally improves estimation efficiency compared to $\dd{Y}$.
Common approaches to adjust for covariate imbalance include post-stratification for discrete covariates~\cite{miratrix2013adjusting} and regression adjustment for general covariates~\cite{lin2013agnostic}.
Similarly, when applying CUPED estimators, we should always adjust for pre-experiment covariate imbalance, either by using an augmentation of the form $\Delta(f(\bX))$, or by first residualizing the outcomes and then applying CUPED to residualized $Y_i^* = Y_i - \EE\left[Y_i|\bX_i\right]$. 

After adjusting for pre-experiment covariates, we consider whether to directly balance on certain covariates to further improve the efficiency of a CUPED estimator. 
Recall
$
\that_{cuped.basic} = \dd{Y} - \that_{0},
$
with
$
\var(\that_{cuped.basic}) = \var(\dd{Y}) + \var(\that_{0}) - 2\cov(\dd{Y}, \that_{0}) .
$
We reduce the variance of $\that_{cuped.basic}$ by removing noise shared between $\Delta(Y)$ and $\that_{0}$. 
In other words, we aim to \emph{maximize} the variation captured in $\that_{0}$ that correlates with variation in $\Delta(Y)$. 
Similarly, from Eq~\eqref{eq:mutrig-var}, we see that to reduce the sampling variance of $\that_{trg1}$, we want to keep as much residual variation due to covariate imbalances as possible,
and only minimally balance covariates in order for the augmentation term to have a mean of zero.
Specifically, under a weak ignorability assumption, if we know ground truth triggering probabilities, or equivalently, the propensity score of a data point from $T0 \cup C$ belonging to $T0$, then using these probabilities to construct weights for $\that_0$~\eqref{eq:h_sn} will lead to the largest variance reduction while keeping $\widehat{\tau}_{trg1}$ unbiased. 
Any additional covariate balancing would lessen the amount of variance reduced.
This follows from the propensity score $e(\bX):=\PP(T0|\bX, T0 \cup C)$ being the coarsest balancing score~\citep{rosenbaum1983central}. 

So, should we balance on a covariate or not? 
For pre-experiment covariates, if regression adjustment has already been applied, then there is no need to further balance on these covariates.
This is because the adjusted $\Delta(Y^*)$ already seeks to exclude noise from $\bX$, so keeping extra variation from $\bX$ in the augmentation $\that_0$ no longer contributes to increased correlation between $\Delta(Y^*)$ and $\that_0$.
If regression adjustment has not been applied, then we should either perform the adjustment, or remove the pre-experiment covariate imbalance via augmentation balancing.

For in-experiment covariates, the default is to not balance.
However, balancing might be needed if there is detectable bias in $\that_0$ when pre-experiment covariates used for adjustment fail to include all confounders of triggering status and outcome.
This additional balancing comes at the cost of reduced efficiency gains.

To summarize, we have a knob to control the tradeoff of bias vs. variance when we include more in-experiment data for balancing. 
On one end, we only use pre-experiment covariates, with the possibility of missing some confounders of triggering status and outcome.
Then $\EE \left[\that_0\right] \neq 0$ and the CUPED estimator is biased, though we will see efficiency gains.
On the other end, we balance on the in-experiment outcome $Y$ and the augmentation $\that_0$ becomes trivially mean-zero. 
Then, the CUPED estimator is unbiased, but provides no efficiency gains. 
The best choice of in-experiment covariates is a combination of science and art that relies on practitioners' domain knowledge.

\section{A Real Experiment}\label{sec:realexp}
We applied our method to a real digital experiment run on 2.5 million users, where 50\% of users were randomly assigned to the treatment group and the rest were assigned to control.
Users in the treatment group were presented with a new feature that only took effect if a user chose to activate the feature. 
Only 5\% of treatment users activated the new feature. 
The target outcome is the count of conversions. 

\begin{table}[htb]
\resizebox{\columnwidth}{!}{%
\begin{tabular}{@{}l|l|l|l@{}}
\toprule
Estimator                             & Variance & Var Reduction & Mean-Zero Test p-value \\ \midrule
Naive Difference-in-Means (baseline)                & 4.65E-4  &           &      N/A               \\
CUPED (regression adjustment)          & 0.55E-4  & 88.2\% (8.5x)     &      N/A                  \\
CUPED OneSidedTrigger (prediction) & 0.03E-4         & 99.2\% (125x)          &  1E-34             \\        
CUPED OneSidedTrigger (balancing in-exp)  & 0.08E-4  & 98.3\% (58.8x)     & 0.485                  \\ \bottomrule
\end{tabular}%
}
\caption{Variance, Variance Reduction, and Mean-Zero Test for various estimators. 
The CUPED estimators here all build upon outcome model adjustment and use residualized $Y$ as the target metric. The balancing approach uses in-experiment features to pass the mean-zero test while sacrificing approximately $2x$ in efficiency gains.}
\label{tab:realexp}
\end{table}

We used 31 pre-experiment features to build two models using XGBoost\cite{Chen:2016xgboost}: 
One model is an outcome regression to predict conversions. 
The other is a triggering probability prediction model.
These two models are then used to construct various estimators of the new feature's overall ITT effect on conversions.
Table~\ref{tab:realexp} summarizes the results. 
Using outcome regression modeling reduced variance by a factor of 8.5, or 88.2\%. 
This alone was very material. 
However, given the low triggering rate of 5\%, we anticipated that a successful application of the one-sided trigger estimator could yield significant efficiency gains beyond standard regression adjustment. 
We first used the predicted triggering probabilities to define weights $w_i$, using the prediction approach described in Section~\ref{sec:weights}. 
The resulting one-sided trigger estimator achieved $125x$ or 99.2\% variance reduction compared to a naive difference-in-means baseline. 
However, the mean-zero test failed with a highly significant p-value of $1\text{E}-34$, which meant we were missing some important confounders of triggering status and outcome. 
Next, following the balancing approach described in Sections~\ref{sec:weights} and~\ref{sec:adjust-or-not}, we introduced two additional in-experiment covariates that measure user activity level during the experiment. 
We built a propensity score model of $e(\bX) := \PP(T0|\bX, T0 \cup C)$ using logistic regression, fit on the original triggering probability predictions and the two in-experiment user activity features.
The predicted triggering probabilities provide a sufficient dimension reduction of the 31 pre-experiment features, so we do not include all the pre-experiment features individually.
We then constructed weights $w_i = \hat{e}_i/(1-\hat{e}_i)$ for the Eq~\eqref{eq:h_sn} augmentation term. 
This balancing approach passed the mean-zero test with a p-value of 0.485. 
Although introducing in-experiment features reduced the efficiency gain ($\approx2x$ loss compared to the prediction approach), this version of mean-zero augmentation still achieved a $58.8x$ variance reduction (98.3\%) compared to baseline, which is another $7x$ gain on top of CUPED with outcome modeling regression adjustment. 

While the prediction and balancing approaches behind the One-Sided Trigger Estimator share many theoretical foundations with observational causal inference techniques, a randomized design paired with a testable mean-zero assumption make our estimation approach more defensible in practice. 
In our experience, it is often difficult to pass the mean-zero test on the first try.
For real applications, applying the one-sided triggering technique likely involves an iterative cycle of failing the mean-zero test and adding pre-experiment or in-experiment covariates until the mean-zero augmentation condition is satisfied.
These iterative checks can largely be automated, similar to tools that use hypothesis testing to do variable selection.

\section{Simulation Study}\label{sec:simulation}
We ran four simulation studies to validate and strengthen the understanding of our method. 
In the first study, we implement the One-Sided Trigger as described in Section~\ref{sec:oneside-trigger-steps} and compare its average bias and standard error against that of the Naive estimator $\dd{Y}$. 
We also compare to estimators that could be used when triggering status is fully observed in both treatment and control groups.
In Study 2, we compare the prediction approach to the balancing approach for setting weights used in the CUPED augmentation term (ref. Section~\ref{sec:weights}). 
We demonstrate the phenomenon of increased variance due to covariate overbalancing when we use in-sample propensity score predictions (ref. Section~\ref{sec:cov_bal}). 
Study 3 shows the impact of combining regression adjustment with one-sided triggering augmentation. 
Study 4 provides an example where observed pre-experiment covariates are insufficient to make the augmentation mean-zero, and it is necessary to balance on an in-experiment covariate. 
We share the code\footnote{\url{https://osf.io/kum6d/?view_only=8c2f0fe6c40c40029faecb7a583f0c45}} to replicate the simulation study.

\subsection{Simulation Setup and Evaluation Methods}\label{sec:dgp}
\begin{figure}
    \centering
    \includegraphics[clip, trim=0.01cm 1.2cm 0.1cm 1.3cm, width=0.6\columnwidth]{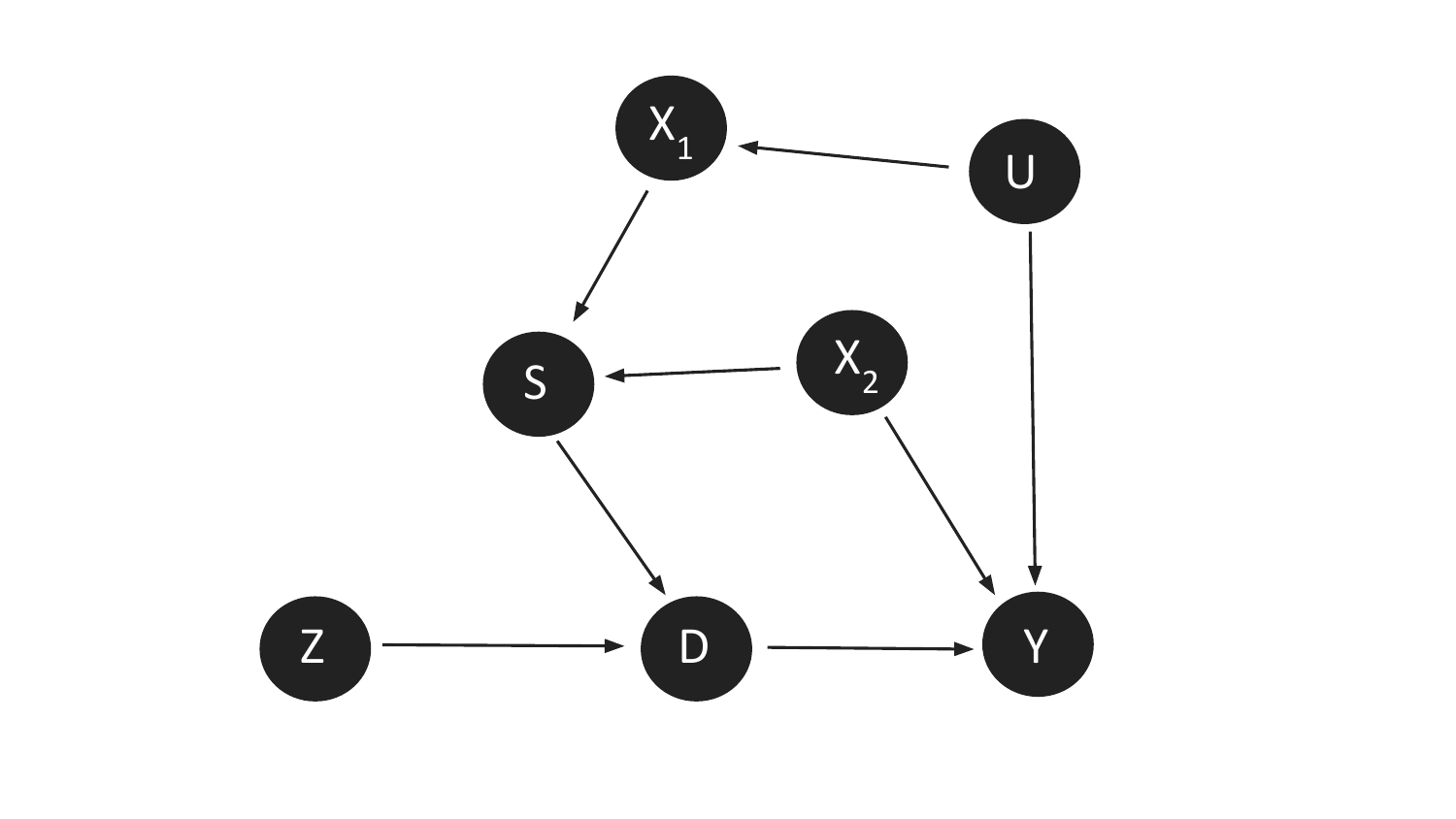}
    \caption{DAG for the simulation design. $Z$ is treatment assignment, $D$ is the actual triggering event, and $Y$ is the outcome. $S$ is the latent triggering counterfactual status. $X_1$ and $X_2$ are pre-experiment covariates that are confounders of $S$ and $Y$. $U$ is an unobserved confounder, but $X_1$ and $X_2$ fully block all backdoor paths from $S$ to $Y$.}
    \label{fig:dag}
\end{figure}

Our simulation design (Figure~\ref{fig:dag}) mimics a conversion process often of interest in online experiments.
We outline the data generating process (dgp) and refer the reader to the companion code for details.

We simulate a randomized experiment with $n=100,000$ users, of which $n_T = 75,000$ are assigned to treatment and the remaining $n_C = 25,000$ are assigned to control. 
Each user $i$ belongs to either a high or low engagement tier, represented by an unobserved binary label $U_i$. 
Two pre-experiment covariates, $X_{1i}$ and $X_{2i}$, are generated from uniform distributions with a lower bound of 0.
The upper-bound for $X_{1i}$ is $1$ when $U_i=1$, and $0.25$ when $U_i=0$;  
this allows us to set a higher conversion rate for high-engagement users.
The upper bound for $X_{2i}$ is $1$.
The user-specific triggering rate (which parameterizes the triggering counterfactual status $S_i$) is a linear function of $X_{1i}$ and $X_{2i}$. 
The user-specific baseline conversion rate is a linear function of $X_{2i}$ and $U_i$.
For those who triggered in the treatment group, we add an additional constant treatment effect to the conversion rate. 
The outcome $Y_i$ is generated from a binomial distribution with $30$ trials and probability of success equal to the user-specific conversion rate, which can be interpreted as total conversions in a 30-day period under a fixed daily conversion rate. 
From Figure~\ref{fig:dag}, we see that $U$, $X_1$, and $X_2$ are confounders that affect both the triggering status $S$ and the outcome $Y$.
The weak principal ignorability Assumption~\eqref{ass:ignorability} is satisfied when we condition on $X_1$ and $X_2$, thereby blocking all backdoor paths from $S$ to $Y$~\cite{pearl2000models}. 
Unless otherwise stated, we assume $S$ is only observed in the treatment group, i.e., we have an experiment with one-sided triggering.

\noindent The most important statistics for this simulation study are as follows:
\begin{itemize}
\setlength{\itemindent}{1em}
    \item The ground truth ITT effect is \emph{0.075}
    \item The triggering rate is \emph{5\%}
    \item $n_T = 75,000$ and $n_C = 25,000$
\end{itemize}

\noindent For each study, we run 50,000 simulation trials and report the following evaluations for each estimator: 
\begin{itemize}
\setlength{\itemindent}{1em}
    \item \emph{Est. ITT}. Estimated ITT. We compare this to the true ITT effect of 0.075 to assess the \emph{mean bias} of the estimator.
    \item \emph{True SE}. Sample standard deviation across the 50,000 trials. We use this as the estimator's true (Monte Carlo) standard error to compare \emph{sample efficiency}.
    \item \emph{Est. SE}. Average estimated standard error. We estimate the SE of all One-Sided Trigger estimators with 1000 bootstrap samples. We estimate SE for the Naive Difference-in-Means and Trigger-Dilute estimators using closed form formulas. We compare this to the Monte Carlo SE to validate that our SE estimation is not biased. 
\end{itemize}


\noindent In the following sections, we summarize results from each study.

\subsection{Study 1: Benchmark against other estimators }\label{sec:study1}
We first benchmark the CUPED One-Sided Trigger against other unbiased ITT estimators including Naive $\Delta(Y)$, Trigger-Dilute, and the CUPED Two-Sided Trigger from Section ~\ref{sec:cuped-twoside}.
To obtain Trigger-Dilute and Two-Sided Trigger estimates, we \emph{pretend} $S$ is observed for control subjects.
We implement the One-Sided Trigger following the prediction approach, where the weights $w_i$ are predicted triggering probabilities $\PP(S_i=1|X_{1,i},X_{2,i})$. 
In particular, for each simulation trial, we fit a logistic regression $S\sim X_1+X_2$ using treatment group data and make (out-of-sample) predictions on the control group.

Simulation results are in Table~\ref{tab:simulation1}. We make several observations.
First, all estimators are unbiased for the true ITT of 0.075. 
Second, Trigger-Dilute and Two-Sided Trigger have the same ground truth variance and both reduce SE by about 4 times compared to Naive. 
This is roughly a 16x variance reduction rate, and corroborates the heuristic that variance reduction with trigger-dilute analysis can be as high as the reciprocal of the triggering rate (which is set at $5\%$ in our dgp). 
The proposed One-Sided Trigger estimator has the smallest SE. 
This means that when the weak principal ignorability assumption holds, not observing triggering status $S$ in the control group does not prevent us from using triggering to reduce variance. 
In fact, exploiting the ignorability assumption allows us to obtain precision improvements even beyond Trigger-Dilute. 
Third, all the variance estimators indeed recover the ground truth SE.
In particular, we have confirmed that the bootstrap procedure described in~\ref{sec:cuped-oneside} is able to recover the true SE of the One-Sided Trigger estimator. 

\begin{table}[htb]
\resizebox{0.7\columnwidth}{!}{%
\begin{tabular}{@{}l|l|l|l@{}}
\toprule
Estimator                     & Est. ITT & True SE & Est. SE \\ \midrule
Naive                   & 0.0750               & 0.0122                    & 0.0123                              \\
Trigger-Dilute                & 0.0750               & 0.00315                   & 0.00315   \\
Two-Sided Trigger       & 0.0750               & 0.00315                   & 0.00324                             \\
One-Sided Trigger       & 0.0750               & 0.00195                   & 0.00195           \\ \bottomrule
\end{tabular}%
}
\caption{(Study 1) Performance of One-Sided Trigger compared to Naive and two other unbiased estimators that could be used if triggering counterfactual status were fully observed for all units.}
\label{tab:simulation1}
\end{table}

It is surprising that the One-Sided Trigger outperforms both Trigger-Dilute and Two-Sided Trigger, even though the latter two estimators have the benefit of observing $S$ for both control and treatment groups.
The extra efficiency gain for the One-Sided Trigger likely comes from its exploitation of weak principal ignorability, which allows it to use the entire control group (i.e., a larger sample size) to obtain a more precise estimate of the control outcome mean for the $C0$ group.
In our simulation setup, only $25\%$ of units are assigned to control, so increasing the effective sample size of the control group can help reduce the variance of estimated control means, which are subsequently used to estimate the average treatment effect.

\subsection{Study 2: Prediction vs. Balancing approach}
This study explores different ways to balance $C$ and $T0$ to create an augmentation term for the One-Sided Trigger.
Specifically, we compare the prediction and balancing approaches for finding weights in Eq~\eqref{eq:h_sn}.
Results are shown in Table~\ref{tab:simulation2}.
All the One-Sided Trigger estimators are unbiased, because we correctly account for the confounders $X_1$ and $X_2$ when fitting the triggering probability model and the propensity score model, and in entropy balancing. 
In terms of SE, weights based on the triggering probability prediction approach give the largest variance reduction, with a SE of $0.00195$, which is more than 3 times smaller than the SEs of the balancing-based estimators (SE $\approx 0.00738$).
Among balancing approaches, propensity score and entropy balancing have similar efficiency, although adding an unnecessary covariate $U$ slightly reduces efficiency. 
\ 

\begin{table}[hb]
\resizebox{0.9\columnwidth}{!}{%
\begin{tabular}{@{}l|l|l|l@{}}
\toprule
Estimator                     & Est. ITT & True SE & Est. SE\\ \midrule
Naive                   & 0.0750               & 0.0122                    & 0.0123                              \\
Trigger-Dilute              & 0.0750               & 0.00315                   & 0.00315  \\
\midrule
CUPED One-Sided Trigger estimators: \\
Triggering probability prediction  & 0.0750               & 0.00195                   &      0.00195    \\ 
Triggering probability ground truth  & 0.0750               & 0.00227                  &     0.00228        \\ 
Balancing with propensity score & 0.0750               & 0.00738                   &     0.00738       \\ 
Entropy balancing on $X_1, X_2$  & 0.0750               & 0.00738                   &      0.00737      \\ 
Entropy balancing on $X_1, X_2, U$ & 0.0750               & 0.00797                  &         0.00797   \\ 
\bottomrule
\end{tabular}%
}
\caption{(Study 2) Estimation performance of CUPED One-Sided Trigger under different specifications of the augmentation term.}
\label{tab:simulation2}
\end{table}
\
This study shows that a prediction approach results in larger variance reduction than a balancing approach. 
This is because the predicted triggering probabilities are out-of-sample predictions, whereas the estimated propensity scores are in-sample predictions.
(The triggering probability model is trained on all treatment units, and used to make out-of-sample predictions for control units. The propensity score model is trained on all control units and trigger-complement units in the treated group, $T0 \cup C$, and used to make in-sample predictions of whether a unit belongs to $C$.)
Using in-sample predictions to construct the CUPED augmentation term means there will be less ``good'' variation retained in the augmentation (ref. Section~\ref{sec:cov_bal}) that is correlated with $\dd{Y}$. 
In turn, this means the amount of variance reduced will be smaller.
Similarly, when we entropy balance on $(X_1, X_2, U)$ instead of just $(X_1, X_2)$, we are overbalancing, reducing ``good'' variation, and again see a reduction in efficiency gains.

\subsection{Study 3: Balancing in addition to regression adjustment}\label{sec:study3}
In practice, we always use regression adjustment of pre-experiment covariates.
In this study, we first fit a regression model predicting $Y$ using $(X_1,X_2)$, and then apply our various estimators on residual outcomes $Y^* := Y - \EE\left[Y|X_1,X_2\right]$. 
Results in Table~\ref{tab:simulation3} show that all estimators remain unbiased. 
Both Naive and Trigger-Dilute now have smaller SEs compared to Studies 1 and 2 because of the $(X_1, X_2)$ regression adjustment.
All the CUPED One-Sided Trigger estimators also have smaller SEs compared to Table~\ref{tab:simulation2}. 

This study illustrates various remarks from Section~\ref{sec:cov_bal}.
When we combine regression adjustment with one-sided trigger augmentation, balancing on the same pre-experiment covariates used for regression adjustment \emph{does not} penalize us with increased variance.
In fact, as Table~\ref{tab:simulation3} shows, all the CUPED One-Sided Trigger estimators now have the same standard error, with the exception of the estimator that includes $U$ in entropy balancing.
This exception is not surprising, and illustrates the point that overbalancing on additional covariates that are not included in regression adjustment \emph{will} increase variance. 
In our simulation model (Fig.~\ref{fig:dag}), we know controlling for $(X_1, X_2)$ already satisfies the weak principal ignorability assumption. 
It is unnecessary to further balance on $U$, and doing so will in fact be overbalancing.
Balancing on auxiliary covariates that are not needed to ensure a mean-zero $\that_0$ will reduce the amount of ``good'' covariate imbalance retained in the augmentation term and consequently diminish efficiency gains. 
In real applications, we can always include more covariates in the balancing approach to minimize the risk of missing confounders, in exchange for potentially reduced efficiency gains.


\begin{table}[!hbt]
\resizebox{0.9\columnwidth}{!}{%
\begin{tabular}{@{}l|l|l|l@{}}
\toprule
Estimator                     & Est. ITT & True SE & Est. SE \\ \midrule
Naive                   & 0.0750               & 0.00995                    & 0.00999                              \\
Trigger-Dilute              & 0.0750               & 0.00275                   & 0.00275  \\
\midrule
CUPED One-Sided Trigger Estimators: \\
Triggering probability prediction & 0.0750               & 0.00194                   &    0.00194      \\ 
Triggering probability ground truth   & 0.0750               & 0.00194                  &       0.00194      \\ 
Balancing with propensity score  & 0.0750               & 0.00194                   &       0.00194     \\ 
Entropy balancing on $X_1,X_2$  & 0.0750               & 0.00194                   &       0.00195     \\ 
Entropy balancing on $X_1,X_2, U$ & 0.0750               & 0.00359                  &        0.00357    \\ 
\bottomrule
\end{tabular}%
}
\caption{(Study 3) Estimation performance of various estimators, applied on residual outcomes $Y^* = Y - \EE\left[Y | X_1, X_2\right]$. }
\label{tab:simulation3}
\end{table}

\subsection{Study 4: Use in-experiment covariate for bias-variance tradeoff}
This study explores how One-Sided Trigger estimators perform when weak principal ignorability~\eqref{ass:ignorability} does not hold.
Specifically, we pretend the triggering-impact confounder $X_1$ is not observed and that we only have access to $X_2$ as a pre-experiment covariate. 
$U$, which is also a triggering-impact confounder, is treated as an in-experiment observation, so it cannot be used to fit a triggering probability model, but can be used for covariate balancing to create a CUPED augmentation term. 
This situation is akin to the real experiment we shared in Section~\ref{sec:realexp}.
We replicate the simulation as in Study 3, with the only change that $X_1$ is removed from regression adjustment as well as all model fitting and covariate balancing.

Comparing Table~\ref{tab:simulation4} to Table~\ref{tab:simulation3} shows that Naive and Trigger-Dilute estimators are still unbiased, though now with slightly larger SEs due to missing $X_1$ in the regression adjustment. 
The main difference is that One-Sided Trigger estimators with triggering probability prediction, balancing with propensity score weights, and entropy balancing on $X_2$, are no longer unbiased. 
This confirms that $X_2$ alone does not capture all confounding between triggering status $S$ and outcome $Y$. 
When Assumption~\ref{ass:ignorability} is violated and no further adjustments are applied, the CUPED augmentation term~\eqref{eq:h_sn} will be significantly different from zero, and One-Sided Trigger estimates will be biased. 
(Recall that the real experiment in Section~\ref{sec:realexp} had an extremely low p-value when we did not include in-experiment covariates.)   
Because in our dgp $(X_2, U)$ also satisfies weak principal ignorability (by blocking all backdoor paths from $S$ to $Y$), including the in-experiment covariate $U$ in entropy balancing successfully removes the mean bias, but results in a larger SE compared to its biased counterparts.

\begin{table}[!hbt]
\resizebox{0.9\columnwidth}{!}{%
\begin{tabular}{@{}l|l|l|l@{}}
\toprule
Estimator                     & Est. ITT & True SE & Est. SE \\ \midrule
Naive                   & 0.0750               & 0.0104                   & 0.0105                               \\
Trigger-Dilute              & 0.0750               &      0.00287              & 0.00287  \\
\midrule
CUPED One-Sided Trigger Estimators: \\
Triggering probability prediction & 0.0797               &    0.00203                &   0.00202       \\ 
Balancing with propensity score   & 0.0797               &      0.00203              &     0.00202       \\ 
Entropy balancing on $X_2$  & 0.0797               &         0.00203           &       0.00203     \\ 
Entropy balancing on $X_2, U$ & 0.0750               &        0.00478           &      0.00480      \\ 
\bottomrule
\end{tabular}%
}
\caption{(Study 4) Not observing confounder $X_1$ results in biased CUPED estimators until we introduce an in-experiment covariate $U$ that is correlated with both the triggering counterfactual and outcome. All estimation approaches here are applied on residual outcomes $Y^{**} = Y - \EE\left[Y|X_2\right]$.}
\label{tab:simulation4}
\end{table}

\section{Conclusion and Limitations}\label{sec:conclusion}
When the triggering rate in a randomized experiment is low, it is critical to exploit subjects' triggering counterfactual status to efficiently estimate the overall ITT effect. 
However, it is not always possible to know whether a control subject would have triggered the active treatment had they been assigned to the treatment group. 
This kind of one-sided triggering problem poses a challenge both in theory and in practice. 

This paper tackles one-sided triggering purely as a variance reduction problem.
We reduce the sampling variance of the inefficient difference-in-outcome-means estimator by appending a mean-zero augmentation term, which serves as a form of covariate adjustment.
We derive this mean-zero augmentation by comparing the treated trigger-complement group $T0$ against the entire control group $C$. 
Specifically, we reweight subjects in $C$ to make the distribution of outcome $Y$ among $C$ comparable to that among $T0$. 
It is known in the principal stratification literature that such weights exist and can be estimated as a function of pre-experiment covariates when weak principal ignorability holds. 
Following this theory, we propose our augmentation Eq~\eqref{eq:h_sn} with weights found using either a prediction approach or a balancing approach. 
The latter has the benefit of allowing in-experiment covariates to be included in balancing, in stark contrast to typical covariate balancing applications.

A simulation study shows that our proposed estimator can be \emph{even more efficient} than standard estimators that are used when the triggering counterfactual status is observed in both treatment and control groups. 
Both the prediction and balancing approaches to find control unit weights can result in similar efficiency gains when used in conjunction with covariate regression adjustment from an outcome prediction model. 

In real case studies, as shown in Section~\ref{sec:realexp}, conditioning on pre-experiment covariates may be insufficient to satisfy weak principal ignorability, and our augmentation will fail to pass a mean-zero test. For such scenarios, we find that including in-experiment observations in the balancing approach can effectively reduce or eliminate bias, in exchange for extra variance. 
In the worst case, we balance on the in-experiment outcome-of-interest so the augmentation becomes a point mass at zero and we forfeit any efficiency gains. 
However, in practice there are often plenty of available in-experiment observations correlated with the triggering counterfactual, such that controlling for or balancing on these covariates can mitigate confounding between triggering status and the outcome. 
This sets our method apart from traditional balancing procedures in the observational studies literature, where it is typically forbidden to use in-experiment data for adjustment.

Our variance reduction method relies on the construction of an augmentation term that equals zero in expectation.
Although this condition can be tested, the mean-zero test has limited statistical power for small biases, so it does not guarantee complete removal of mean bias.
Experimenters need to trade off potential undetected bias against precision gains.
When an ITT estimate is inconclusive and unhelpful for making business decisions, it may be worthwhile to sacrifice small levels of unbiasedness for large efficiency gains that will support decision-making and product iterations.
  
\clearpage
\bibliographystyle{ACM-Reference-Format}
\balance
\bibliography{sigproc} 


\begin{thebibliography}{42}


\ifx \showCODEN    \undefined \def \showCODEN     #1{\unskip}     \fi
\ifx \showDOI      \undefined \def \showDOI       #1{#1}\fi
\ifx \showISBNx    \undefined \def \showISBNx     #1{\unskip}     \fi
\ifx \showISBNxiii \undefined \def \showISBNxiii  #1{\unskip}     \fi
\ifx \showISSN     \undefined \def \showISSN      #1{\unskip}     \fi
\ifx \showLCCN     \undefined \def \showLCCN      #1{\unskip}     \fi
\ifx \shownote     \undefined \def \shownote      #1{#1}          \fi
\ifx \showarticletitle \undefined \def \showarticletitle #1{#1}   \fi
\ifx \showURL      \undefined \def \showURL       {\relax}        \fi
\providecommand\bibfield[2]{#2}
\providecommand\bibinfo[2]{#2}
\providecommand\natexlab[1]{#1}
\providecommand\showeprint[2][]{arXiv:#2}

\bibitem[Angrist et~al\mbox{.}(1996)]%
        {angrist1996}
\bibfield{author}{\bibinfo{person}{Joshua~D. Angrist},
  \bibinfo{person}{Guido~W. Imbens}, {and} \bibinfo{person}{Donald~B. Rubin}.}
  \bibinfo{year}{1996}\natexlab{}.
\newblock \showarticletitle{Identification of Causal Effects Using Instrumental
  Variables}.
\newblock \bibinfo{journal}{\emph{J. Amer. Statist. Assoc.}}
  \bibinfo{volume}{91}, \bibinfo{number}{434} (\bibinfo{year}{1996}),
  \bibinfo{pages}{444--455}.
\newblock


\bibitem[Bakshy et~al\mbox{.}(2014)]%
        {bakshy2014designing}
\bibfield{author}{\bibinfo{person}{Eytan Bakshy}, \bibinfo{person}{Dean
  Eckles}, {and} \bibinfo{person}{Michael~S Bernstein}.}
  \bibinfo{year}{2014}\natexlab{}.
\newblock \showarticletitle{Designing and deploying online field experiments}.
  In \bibinfo{booktitle}{\emph{Proceedings of the 23rd international conference
  on World wide web}}. \bibinfo{pages}{283--292}.
\newblock


\bibitem[Chattopadhyay and Zubizarreta(2022)]%
        {chattopandhyay2022}
\bibfield{author}{\bibinfo{person}{Ambarish Chattopadhyay} {and}
  \bibinfo{person}{José~R Zubizarreta}.} \bibinfo{year}{2022}\natexlab{}.
\newblock \showarticletitle{{On the implied weights of linear regression for
  causal inference}}.
\newblock \bibinfo{journal}{\emph{Biometrika}} (\bibinfo{date}{10}
  \bibinfo{year}{2022}).
\newblock
\showISSN{1464-3510}
\urldef\tempurl%
\url{https://doi.org/10.1093/biomet/asac058}
\showDOI{\tempurl}
\showeprint{https://academic.oup.com/biomet/advance-article-pdf/doi/10.1093/biomet/asac058/46684037/asac058.pdf}
\newblock
\shownote{asac058}.


\bibitem[Chen and Guestrin(2016)]%
        {Chen:2016xgboost}
\bibfield{author}{\bibinfo{person}{Tianqi Chen} {and} \bibinfo{person}{Carlos
  Guestrin}.} \bibinfo{year}{2016}\natexlab{}.
\newblock \showarticletitle{{XGBoost}: A Scalable Tree Boosting System}. In
  \bibinfo{booktitle}{\emph{Proceedings of the 22nd ACM SIGKDD International
  Conference on Knowledge Discovery and Data Mining}} (San Francisco,
  California, USA) \emph{(\bibinfo{series}{KDD '16})}.
  \bibinfo{publisher}{ACM}, \bibinfo{address}{New York, NY, USA},
  \bibinfo{pages}{785--794}.
\newblock
\showISBNx{978-1-4503-4232-2}
\urldef\tempurl%
\url{https://doi.org/10.1145/2939672.2939785}
\showDOI{\tempurl}


\bibitem[Coussens and Spiess(2021)]%
        {Coussens2021ImprovingIF}
\bibfield{author}{\bibinfo{person}{Stephen Coussens} {and}
  \bibinfo{person}{Jann Spiess}.} \bibinfo{year}{2021}\natexlab{}.
\newblock \showarticletitle{Improving Inference from Simple Instruments through
  Compliance Estimation}.
\newblock


\bibitem[Deng and Hu(2015)]%
        {deng2015diluted}
\bibfield{author}{\bibinfo{person}{Alex Deng} {and} \bibinfo{person}{Victor
  Hu}.} \bibinfo{year}{2015}\natexlab{}.
\newblock \showarticletitle{Diluted treatment effect estimation for trigger
  analysis in online controlled experiments}. In
  \bibinfo{booktitle}{\emph{Proceedings of the Eighth ACM International
  Conference on Web Search and Data Mining}}. \bibinfo{pages}{349--358}.
\newblock


\bibitem[Deng et~al\mbox{.}(2018)]%
        {Dengkdd2018}
\bibfield{author}{\bibinfo{person}{Alex Deng}, \bibinfo{person}{Ulf Knoblich},
  {and} \bibinfo{person}{Jiannan Lu}.} \bibinfo{year}{2018}\natexlab{}.
\newblock \showarticletitle{Applying the Delta Method in Metric Analytics: A
  Practical Guide with Novel Ideas}. In \bibinfo{booktitle}{\emph{Proceedings
  of the 24th ACM SIGKDD International Conference on Knowledge Discovery and
  Data Mining}} (London, United Kingdom) \emph{(\bibinfo{series}{KDD '18})}.
  \bibinfo{publisher}{ACM}, \bibinfo{address}{New York, NY, USA},
  \bibinfo{pages}{233--242}.
\newblock
\showISBNx{978-1-4503-5552-0}


\bibitem[Deng and Shi(2016)]%
        {Deng:2016b}
\bibfield{author}{\bibinfo{person}{Alex Deng} {and} \bibinfo{person}{Xiaolin
  Shi}.} \bibinfo{year}{2016}\natexlab{}.
\newblock \showarticletitle{Data-driven metric development for online
  controlled experiments: Seven lessons learned}. In
  \bibinfo{booktitle}{\emph{Proceedings of the 22nd ACM SIGKDD International
  Conference on Knowledge Discovery and Data Mining}}.
\newblock


\bibitem[Deng et~al\mbox{.}(2013)]%
        {deng2013cuped}
\bibfield{author}{\bibinfo{person}{Alex Deng}, \bibinfo{person}{Ya Xu},
  \bibinfo{person}{Ron Kohavi}, {and} \bibinfo{person}{Toby Walker}.}
  \bibinfo{year}{2013}\natexlab{}.
\newblock \showarticletitle{Improving the sensitivity of online controlled
  experiments by utilizing pre-experiment data}. In
  \bibinfo{booktitle}{\emph{Proceedings of the 6th ACM WSDM Conference}}.
  \bibinfo{pages}{123--132}.
\newblock


\bibitem[Ding and Lu(2017)]%
        {ding2017principal}
\bibfield{author}{\bibinfo{person}{Peng Ding} {and} \bibinfo{person}{Jiannan
  Lu}.} \bibinfo{year}{2017}\natexlab{}.
\newblock \showarticletitle{Principal stratification analysis using principal
  scores}.
\newblock \bibinfo{journal}{\emph{Journal of the Royal Statistical Society:
  Series B (Statistical Methodology)}} \bibinfo{volume}{79},
  \bibinfo{number}{3} (\bibinfo{year}{2017}), \bibinfo{pages}{757--777}.
\newblock


\bibitem[Drutsa et~al\mbox{.}(2019)]%
        {drutsa2019effective}
\bibfield{author}{\bibinfo{person}{Alexey Drutsa}, \bibinfo{person}{Gleb
  Gusev}, \bibinfo{person}{Eugene Kharitonov}, \bibinfo{person}{Denis
  Kulemyakin}, \bibinfo{person}{Pavel Serdyukov}, {and} \bibinfo{person}{Igor
  Yashkov}.} \bibinfo{year}{2019}\natexlab{}.
\newblock \showarticletitle{Effective online evaluation for web search}. In
  \bibinfo{booktitle}{\emph{Proceedings of the 42nd International ACM SIGIR
  Conference on Research and Development in Information Retrieval}}.
  \bibinfo{pages}{1399--1400}.
\newblock


\bibitem[Efron and Tibshirani(1994)]%
        {efron1994introduction}
\bibfield{author}{\bibinfo{person}{Bradley Efron} {and}
  \bibinfo{person}{Robert~J Tibshirani}.} \bibinfo{year}{1994}\natexlab{}.
\newblock \bibinfo{booktitle}{\emph{An introduction to the bootstrap}}.
\newblock \bibinfo{publisher}{CRC press}.
\newblock


\bibitem[Fabijan et~al\mbox{.}(2019)]%
        {fabijan2019diagnosing}
\bibfield{author}{\bibinfo{person}{Aleksander Fabijan}, \bibinfo{person}{Jayant
  Gupchup}, \bibinfo{person}{Somit Gupta}, \bibinfo{person}{Jeff Omhover},
  \bibinfo{person}{Wen Qin}, \bibinfo{person}{Lukas Vermeer}, {and}
  \bibinfo{person}{Pavel Dmitriev}.} \bibinfo{year}{2019}\natexlab{}.
\newblock \showarticletitle{Diagnosing Sample Ratio Mismatch in Online
  Controlled Experiments: A Taxonomy and Rules of Thumb for Practitioners}. In
  \bibinfo{booktitle}{\emph{Proceedings of the 25th ACM SIGKDD International
  Conference on Knowledge Discovery \& Data Mining}}. ACM,
  \bibinfo{pages}{2156--2164}.
\newblock


\bibitem[Feller et~al\mbox{.}(2017)]%
        {feller2017}
\bibfield{author}{\bibinfo{person}{Avi Feller}, \bibinfo{person}{Fabrizia
  Mealli}, {and} \bibinfo{person}{Luke Miratrix}.}
  \bibinfo{year}{2017}\natexlab{}.
\newblock \showarticletitle{Principal Score Methods: Assumptions, Extensions,
  and Practical Considerations}.
\newblock \bibinfo{journal}{\emph{Journal of Educational and Behavioral
  Statistics}} \bibinfo{volume}{42}, \bibinfo{number}{6}
  (\bibinfo{year}{2017}), \bibinfo{pages}{726--758}.
\newblock


\bibitem[Fisher(1925)]%
        {fisher1925statistical}
\bibfield{author}{\bibinfo{person}{Sir Ronald~Aylmer Fisher}.}
  \bibinfo{year}{1925}\natexlab{}.
\newblock \bibinfo{booktitle}{\emph{Statistical methods for research workers}}.
  Vol.~\bibinfo{volume}{1}.
\newblock \bibinfo{publisher}{Oliver and Boyd Edinburgh}.
\newblock


\bibitem[Frangakis and Rubin(2002)]%
        {frangakis2002}
\bibfield{author}{\bibinfo{person}{Constantine~E. Frangakis} {and}
  \bibinfo{person}{Donald~B. Rubin}.} \bibinfo{year}{2002}\natexlab{}.
\newblock \showarticletitle{Principal Stratification in Causal Inference}.
\newblock \bibinfo{journal}{\emph{Biometrics}} \bibinfo{volume}{58},
  \bibinfo{number}{1} (\bibinfo{year}{2002}), \bibinfo{pages}{21--29}.
\newblock


\bibitem[Guo et~al\mbox{.}(2021)]%
        {guo2021machine}
\bibfield{author}{\bibinfo{person}{Yongyi Guo}, \bibinfo{person}{Dominic Coey},
  \bibinfo{person}{Mikael Konutgan}, \bibinfo{person}{Wenting Li},
  \bibinfo{person}{Chris Schoener}, {and} \bibinfo{person}{Matt Goldman}.}
  \bibinfo{year}{2021}\natexlab{}.
\newblock \showarticletitle{Machine Learning for Variance Reduction in Online
  Experiments}.
\newblock \bibinfo{journal}{\emph{arXiv preprint arXiv:2106.07263}}
  (\bibinfo{year}{2021}).
\newblock


\bibitem[Gupta et~al\mbox{.}(2019)]%
        {Gupta:2019}
\bibfield{author}{\bibinfo{person}{Somit Gupta} {et~al\mbox{.}}}
  \bibinfo{year}{2019}\natexlab{}.
\newblock \showarticletitle{Top Challenges from the First Practical Online
  Controlled Experiments Summit}.
\newblock \bibinfo{journal}{\emph{SIGKDD Explor. Newsl.}} \bibinfo{volume}{21},
  \bibinfo{number}{1} (\bibinfo{date}{May} \bibinfo{year}{2019}),
  \bibinfo{pages}{20--35}.
\newblock
\showISSN{1931-0145}


\bibitem[Hainmueller(2012)]%
        {hainmueller2012entropy}
\bibfield{author}{\bibinfo{person}{Jens Hainmueller}.}
  \bibinfo{year}{2012}\natexlab{}.
\newblock \showarticletitle{Entropy balancing for causal effects: A
  multivariate reweighting method to produce balanced samples in observational
  studies}.
\newblock \bibinfo{journal}{\emph{Political analysis}} \bibinfo{volume}{20},
  \bibinfo{number}{1} (\bibinfo{year}{2012}), \bibinfo{pages}{25--46}.
\newblock


\bibitem[Huntington-Klein(2020)]%
        {HuntingtonKlein_2020}
\bibfield{author}{\bibinfo{person}{Nick Huntington-Klein}.}
  \bibinfo{year}{2020}\natexlab{}.
\newblock \showarticletitle{Instruments with Heterogeneous Effects: Bias,
  Monotonicity, and Localness:}.
\newblock \bibinfo{journal}{\emph{Journal of Causal Inference}}
  \bibinfo{volume}{8}, \bibinfo{number}{1} (\bibinfo{year}{2020}),
  \bibinfo{pages}{182--208}.
\newblock
\urldef\tempurl%
\url{https://doi.org/doi:10.1515/jci-2020-0011}
\showDOI{\tempurl}


\bibitem[Imai and Ratkovic(2014)]%
        {imai2014covariate}
\bibfield{author}{\bibinfo{person}{Kosuke Imai} {and} \bibinfo{person}{Marc
  Ratkovic}.} \bibinfo{year}{2014}\natexlab{}.
\newblock \showarticletitle{Covariate balancing propensity score}.
\newblock \bibinfo{journal}{\emph{Journal of the Royal Statistical Society:
  Series B (Statistical Methodology)}} \bibinfo{volume}{76},
  \bibinfo{number}{1} (\bibinfo{year}{2014}), \bibinfo{pages}{243--263}.
\newblock


\bibitem[Imbens and Angrist(1994)]%
        {angrist1994}
\bibfield{author}{\bibinfo{person}{Guido Imbens} {and} \bibinfo{person}{Joshua
  Angrist}.} \bibinfo{year}{1994}\natexlab{}.
\newblock \showarticletitle{Identification and Estimation of Local Average
  Treatment Effects}.
\newblock \bibinfo{journal}{\emph{Econometrica}} \bibinfo{volume}{62},
  \bibinfo{number}{2} (\bibinfo{year}{1994}), \bibinfo{pages}{467--75}.
\newblock


\bibitem[Imbens and Rubin(2015)]%
        {Imbens:2015}
\bibfield{author}{\bibinfo{person}{G.~W. Imbens} {and} \bibinfo{person}{D.~B.
  Rubin}.} \bibinfo{year}{2015}\natexlab{}.
\newblock \bibinfo{booktitle}{\emph{Causal Inference in Statistics, Social, and
  Biomedical Sciences: An Introduction}}.
\newblock \bibinfo{publisher}{New York: Cambridge University Press}.
\newblock


\bibitem[Jiang and Ding(2020)]%
        {jiang2020identification}
\bibfield{author}{\bibinfo{person}{Zhichao Jiang} {and} \bibinfo{person}{Peng
  Ding}.} \bibinfo{year}{2020}\natexlab{}.
\newblock \showarticletitle{Identification of causal effects within principal
  strata using auxiliary variables}.
\newblock \bibinfo{journal}{\emph{arXiv preprint arXiv:2008.02703}}
  (\bibinfo{year}{2020}).
\newblock


\bibitem[Jo and Stuart(2009)]%
        {jo_stuart2009}
\bibfield{author}{\bibinfo{person}{Booil Jo} {and}
  \bibinfo{person}{Elizabeth~A. Stuart}.} \bibinfo{year}{2009}\natexlab{}.
\newblock \showarticletitle{On the use of propensity scores in principal causal
  effect estimation}.
\newblock \bibinfo{journal}{\emph{Statistics in Medicine}}
  \bibinfo{volume}{28}, \bibinfo{number}{23} (\bibinfo{year}{2009}),
  \bibinfo{pages}{2857--2875}.
\newblock


\bibitem[Joffe and Brensinger(2003)]%
        {joffe2003weighting}
\bibfield{author}{\bibinfo{person}{Marshall~M Joffe} {and}
  \bibinfo{person}{Colleen Brensinger}.} \bibinfo{year}{2003}\natexlab{}.
\newblock \showarticletitle{Weighting in instrumental variables and
  G-estimation}.
\newblock \bibinfo{journal}{\emph{Statistics in medicine}}
  \bibinfo{volume}{22}, \bibinfo{number}{8} (\bibinfo{year}{2003}),
  \bibinfo{pages}{1285--1303}.
\newblock


\bibitem[Kohavi et~al\mbox{.}(2013)]%
        {abScale}
\bibfield{author}{\bibinfo{person}{Ron Kohavi}, \bibinfo{person}{Alex Deng},
  \bibinfo{person}{Brian Frasca}, \bibinfo{person}{Toby Walker},
  \bibinfo{person}{Ya Xu}, {and} \bibinfo{person}{Nils Pohlmann}.}
  \bibinfo{year}{2013}\natexlab{}.
\newblock \showarticletitle{Online controlled experiments at large scale}. In
  \bibinfo{booktitle}{\emph{Proceedings of the 19th ACM SIGKDD international
  conference on Knowledge discovery and data mining}}.
  \bibinfo{pages}{1168--1176}.
\newblock


\bibitem[Kohavi et~al\mbox{.}(2007)]%
        {kohavi2007practical}
\bibfield{author}{\bibinfo{person}{Ron Kohavi}, \bibinfo{person}{Randal~M
  Henne}, {and} \bibinfo{person}{Dan Sommerfield}.}
  \bibinfo{year}{2007}\natexlab{}.
\newblock \showarticletitle{Practical guide to controlled experiments on the
  web: listen to your customers not to the hippo}. In
  \bibinfo{booktitle}{\emph{Proceedings of the 13th ACM SIGKDD Conference}}.
  \bibinfo{pages}{959--967}.
\newblock


\bibitem[Kohavi et~al\mbox{.}(2009)]%
        {kohavi2009controlled}
\bibfield{author}{\bibinfo{person}{Ron Kohavi}, \bibinfo{person}{Roger
  Longbotham}, \bibinfo{person}{Dan Sommerfield}, {and}
  \bibinfo{person}{Randal~M Henne}.} \bibinfo{year}{2009}\natexlab{}.
\newblock \showarticletitle{Controlled experiments on the web: survey and
  practical guide}.
\newblock \bibinfo{journal}{\emph{Data Mining and Knowledge Discovery}}
  \bibinfo{volume}{18}, \bibinfo{number}{1} (\bibinfo{year}{2009}),
  \bibinfo{pages}{140--181}.
\newblock


\bibitem[Kohavi et~al\mbox{.}(2020)]%
        {kohavi2020trustworthy}
\bibfield{author}{\bibinfo{person}{Ron Kohavi}, \bibinfo{person}{Diane Tang},
  {and} \bibinfo{person}{Ya Xu}.} \bibinfo{year}{2020}\natexlab{}.
\newblock \bibinfo{booktitle}{\emph{Trustworthy online controlled experiments:
  A practical guide to a/b testing}}.
\newblock \bibinfo{publisher}{Cambridge University Press}.
\newblock


\bibitem[Li and Ding(2020)]%
        {liding2020}
\bibfield{author}{\bibinfo{person}{Xinran Li} {and} \bibinfo{person}{Peng
  Ding}.} \bibinfo{year}{2020}\natexlab{}.
\newblock \showarticletitle{Rerandomization and regression adjustment}.
\newblock \bibinfo{journal}{\emph{Journal of the Royal Statistical Society:
  Series B (Statistical Methodology)}} \bibinfo{volume}{82},
  \bibinfo{number}{1} (\bibinfo{year}{2020}), \bibinfo{pages}{241--268}.
\newblock
\urldef\tempurl%
\url{https://doi.org/10.1111/rssb.12353}
\showDOI{\tempurl}
\showeprint{https://rss.onlinelibrary.wiley.com/doi/pdf/10.1111/rssb.12353}


\bibitem[Lin(2013)]%
        {lin2013agnostic}
\bibfield{author}{\bibinfo{person}{Winston Lin}.}
  \bibinfo{year}{2013}\natexlab{}.
\newblock \showarticletitle{Agnostic notes on regression adjustments to
  experimental data: Reexamining Freedman’s critique}.
\newblock \bibinfo{journal}{\emph{The Annals of Applied Statistics}}
  \bibinfo{volume}{7}, \bibinfo{number}{1} (\bibinfo{year}{2013}),
  \bibinfo{pages}{295--318}.
\newblock


\bibitem[Miratrix et~al\mbox{.}(2013)]%
        {miratrix2013adjusting}
\bibfield{author}{\bibinfo{person}{Luke~W Miratrix}, \bibinfo{person}{Jasjeet~S
  Sekhon}, {and} \bibinfo{person}{Bin Yu}.} \bibinfo{year}{2013}\natexlab{}.
\newblock \showarticletitle{Adjusting treatment effect estimates by
  post-stratification in randomized experiments}.
\newblock \bibinfo{journal}{\emph{Journal of the Royal Statistical Society:
  Series B (Statistical Methodology)}} \bibinfo{volume}{75},
  \bibinfo{number}{2} (\bibinfo{year}{2013}), \bibinfo{pages}{369--396}.
\newblock


\bibitem[Pearl et~al\mbox{.}(2000)]%
        {pearl2000models}
\bibfield{author}{\bibinfo{person}{Judea Pearl} {et~al\mbox{.}}}
  \bibinfo{year}{2000}\natexlab{}.
\newblock \showarticletitle{Models, reasoning and inference}.
\newblock \bibinfo{journal}{\emph{Cambridge, UK: CambridgeUniversityPress}}
  \bibinfo{volume}{19} (\bibinfo{year}{2000}).
\newblock


\bibitem[Poyarkov et~al\mbox{.}(2016)]%
        {poyarkov2016boosted}
\bibfield{author}{\bibinfo{person}{Alexey Poyarkov}, \bibinfo{person}{Alexey
  Drutsa}, \bibinfo{person}{Andrey Khalyavin}, \bibinfo{person}{Gleb Gusev},
  {and} \bibinfo{person}{Pavel Serdyukov}.} \bibinfo{year}{2016}\natexlab{}.
\newblock \showarticletitle{Boosted decision tree regression adjustment for
  variance reduction in online controlled experiments}. In
  \bibinfo{booktitle}{\emph{Proceedings of the 22nd ACM SIGKDD International
  Conference on Knowledge Discovery and Data Mining}}.
  \bibinfo{pages}{235--244}.
\newblock


\bibitem[Qingyuan and Daniel(2017)]%
        {qingyuan2017entropy}
\bibfield{author}{\bibinfo{person}{Zhao Qingyuan} {and}
  \bibinfo{person}{Percival Daniel}.} \bibinfo{year}{2017}\natexlab{}.
\newblock \showarticletitle{Entropy Balancing is Doubly Robust}.
\newblock \bibinfo{journal}{\emph{Journal of Causal Inference}}
  \bibinfo{volume}{5}, \bibinfo{number}{1} (\bibinfo{year}{2017}).
\newblock


\bibitem[Rosenbaum and Rubin(1983)]%
        {rosenbaum1983central}
\bibfield{author}{\bibinfo{person}{Paul~R Rosenbaum} {and}
  \bibinfo{person}{Donald~B Rubin}.} \bibinfo{year}{1983}\natexlab{}.
\newblock \showarticletitle{The central role of the propensity score in
  observational studies for causal effects}.
\newblock \bibinfo{journal}{\emph{Biometrika}} \bibinfo{volume}{70},
  \bibinfo{number}{1} (\bibinfo{year}{1983}), \bibinfo{pages}{41--55}.
\newblock


\bibitem[Tang et~al\mbox{.}(2010)]%
        {googlesurvey}
\bibfield{author}{\bibinfo{person}{Diane Tang}, \bibinfo{person}{Ashish
  Agarwal}, \bibinfo{person}{Deirdre O'Brien}, {and} \bibinfo{person}{Mike
  Meyer}.} \bibinfo{year}{2010}\natexlab{}.
\newblock \showarticletitle{Overlapping experiment infrastructure: More,
  better, faster experimentation}. In \bibinfo{booktitle}{\emph{Proceedings of
  the 16th ACM SIGKDD international conference on Knowledge discovery and data
  mining}}. \bibinfo{pages}{17--26}.
\newblock


\bibitem[Xie and Aurisset(2016)]%
        {xie2016improving}
\bibfield{author}{\bibinfo{person}{Huizhi Xie} {and} \bibinfo{person}{Juliette
  Aurisset}.} \bibinfo{year}{2016}\natexlab{}.
\newblock \showarticletitle{Improving the sensitivity of online controlled
  experiments: Case studies at netflix}. In
  \bibinfo{booktitle}{\emph{Proceedings of the 22nd ACM SIGKDD International
  Conference on Knowledge Discovery and Data Mining}}. ACM,
  \bibinfo{pages}{645--654}.
\newblock


\bibitem[Xie et~al\mbox{.}(2021)]%
        {xie2021measure}
\bibfield{author}{\bibinfo{person}{Yuxiang Xie}, \bibinfo{person}{Meng Xu},
  \bibinfo{person}{Evan Chow}, {and} \bibinfo{person}{Xiaolin Shi}.}
  \bibinfo{year}{2021}\natexlab{}.
\newblock \showarticletitle{How to Measure Your App: A Couple of Pitfalls and
  Remedies in Measuring App Performance in Online Controlled Experiments}. In
  \bibinfo{booktitle}{\emph{Proceedings of the 14th ACM International
  Conference on Web Search and Data Mining}}. \bibinfo{pages}{949--957}.
\newblock


\bibitem[Xu et~al\mbox{.}(2015)]%
        {xu2015infrastructure}
\bibfield{author}{\bibinfo{person}{Ya Xu}, \bibinfo{person}{Nanyu Chen},
  \bibinfo{person}{Addrian Fernandez}, \bibinfo{person}{Omar Sinno}, {and}
  \bibinfo{person}{Anmol Bhasin}.} \bibinfo{year}{2015}\natexlab{}.
\newblock \showarticletitle{From infrastructure to culture: A/B testing
  challenges in large scale social networks}. In
  \bibinfo{booktitle}{\emph{Proceedings of the 21th ACM SIGKDD International
  Conference on Knowledge Discovery and Data Mining}}. ACM,
  \bibinfo{pages}{2227--2236}.
\newblock


\bibitem[Yuan et~al\mbox{.}(2019)]%
        {yuan2019}
\bibfield{author}{\bibinfo{person}{Lo-Hua Yuan}, \bibinfo{person}{Avi Feller},
  {and} \bibinfo{person}{Luke~W. Miratrix}.} \bibinfo{year}{2019}\natexlab{}.
\newblock \showarticletitle{{Identifying and estimating principal causal
  effects in a multi-site trial of Early College High Schools}}.
\newblock \bibinfo{journal}{\emph{The Annals of Applied Statistics}}
  \bibinfo{volume}{13}, \bibinfo{number}{3} (\bibinfo{year}{2019}),
  \bibinfo{pages}{1348 -- 1369}.
\newblock


\end{thebibliography}

\end{document}